\documentclass[3p,times,twocolumn,preprint]{elsarticle}

\usepackage{ecrc}

\volume{00}

\firstpage{1}

\journalname{Nuclear Physics B Proceedings Supplement}

\runauth{C. Adam et al.}

\jid{nuphbp}

\jnltitlelogo{Nuclear Physics B Proceedings Supplement}

\usepackage{amsthm}
\usepackage{mathrsfs}

\usepackage[figuresright]{rotating}

\newcommand{\Tr}{\textrm{Tr}}
\newcommand{\Lag}{\mathscr{L}}
\newcommand{\MeV}{\textrm{MeV}}
\newcommand{\fm}{\textrm{fm}}

\begin{document}

\begin{frontmatter}



\dochead{}

\title{A unified approach to nuclei: The BPS Skyrme Model}


\author[USC]{C. Adam}
\author[USC]{\underline{C. Naya}}
\author[USC]{J. Sanchez-Guillen}
\author[Leeds]{J.M. Speight}
\author[USC]{R. Vazquez}
\author[Krakow]{A. Wereszczynski}

\address[USC]{Departamento de Física de Partículas, Universidade de Santiago de Compostela, and Instituto Galego de Física de Altas Enerxías (IGFAE), Facultade de Física-Campus Vida, 15782 Santiago de Compostela, Spain}
\address[Leeds]{School of Mathematics, University of Leeds, Leeds LS2 9JT, England}
\address[Krakow]{Institute of Physics, Jagiellonian University, Reymonta 4, Kraków, Poland}

\begin{abstract}
We present a concrete model of a low energy effective field theory of QCD, the well-known Skyrme Model. Specifically, we will work with the BPS submodel in order to describe the binding energies of nuclei. This BPS Skyrme model is characterized by having a saturated bound for the energy proportional to the baryon number of the nuclei. After presenting this classical result, we will proceed with a semi-classical quantization of the coordinates of spin and isospin. Then, with the further inclusion of the Coulomb interaction as well as a small explicit breaking of the isospin symmetry, we finally calculate the binding energies of nuclei, where an excellent agreement has been found for the nuclei with high baryon number. Besides this, we also apply this model to the study of some thermodynamic properties and to neutron stars.
\end{abstract}

\begin{keyword}


\end{keyword}

\end{frontmatter}

% \linenumbers

\section{Introduction}
\label{Intro}

In the early sixties, Tony H. R. Skyrme proposed an effective theory for the description of the low-energy limit of strong interactions where the primary fields are mesons (more concretely pions) \cite{Skyrme1,Skyrme2}. This novel idea is what nowadays we call Skyrme Model and it was supported when in the large $N_c$ limit of QCD it was found that an effective theory of mesons also arises \cite{Witten}. Then, nuclei appear as collective excitations of these fundamental degrees of freedom and are characterized by a topological charge which is identified with the baryon number, ensuring in this way its conservation. Since as a first step we want to study nuclei and nuclear matter, we will consider the simplest case of two flavours (corresponding to the up and down quarks), so the target space where the Skyrme field $U$ lives will be $SU(2)$:

\begin{equation}
x^\mu \to U(x): \quad \mathbb{R}^3 \times \mathbb{R} \to SU(2)  \nonumber
\end{equation}

Then, the Lagrangian proposed by Skyrme, known as Standard Skyrme Model, has two terms:

\begin{equation}
\Lag = \Lag_2 + \Lag_4
\end{equation}

\noindent where

\begin{equation}
\Lag_2 =-\frac{f_{\pi}^2}{4} \; \Tr \; (U^{\dagger} \partial_{\mu} U \; U^{\dagger} \partial^{\mu} U  ),
\end{equation}

\begin{equation}
\Lag_4 =\frac{1}{32 e^2}\; \Tr \; ([U^{\dagger} \partial_{\mu} U,U^{\dagger} \partial_{\nu} U]^2 ).
\end{equation}

\noindent The first term, $\Lag_2$, is the sigma-model term, quadratic in first derivatives and provides the kinetic energy of pions. On the other hand, the $\Lag_4$ term is known as Skyrme term and is quartic in derivatives. The latter is sufficient to avoid the Derrick theorem so stable solutions can exist. As well, it can be seen that a bound on the energy exists, but in this case, solutions do not saturate it.

Since the Skyrme Model is an effective theory, we can add more terms to the Lagrangian. For instance, we can think this Lagrangian as a derivative expansion, so higher powers of derivatives are expected. The first term we can think of is a potential, $\Lag_0$:

\begin{equation}
\Lag_0 = -\mu^2 \mathcal U(U),
\end{equation}

\noindent which may be related to the pion mass.

If we go to higher derivatives, the next term expected will be sextic:

\begin{equation}
\Lag_6 = - \lambda^2 \pi^4 \mathcal B_\mu^2
\end{equation}

\noindent which is the square of the topological current $B^\mu$

\begin{equation} \label{Bmu}
\mathcal B^\mu = \frac{1}{24 \pi^2}\Tr (\epsilon^{\mu \nu \rho \sigma} U^\dagger \partial_\nu U
 U^\dagger \partial_\rho U U^\dagger \partial_\sigma U).
\end{equation}

Asking the Lagrangian to be no more than quadratic in time derivatives, so a standard hamiltonian formulation is possible, these $\Lag_6$ and $\Lag_0$ are the only extra terms allowed, and the generalized Skyrme Model is

\begin{equation}
\Lag = \Lag_2 + \Lag_4 + \Lag_6 + \Lag_0.
\end{equation}

The idea of this generalization resolves a problem of the Standard Skyrme model, where the binding energies we get from it are too large. With this in mind, in section 2 we present the BPS Skyrme Model \cite{ASW} as a first approximation for a model with small contributions from the $\Lag_2$ and $\Lag_4$ terms. Finally, in section 3 we will apply this model to different aspects of the 'nuclear world': study of the binding energies of nuclei, some thermodynamic properties, and a brief account of neutron stars.

%%%%%%%%%%%%%%%%%%%%%%%%%%%%%%%%%%%%%%%%%%%
%%%%%%%%%%%%%%%%%%%%%%%%%%%%%%%%%%%%%%%%%%%

\section{The BPS Skyrme Model}

The BPS Skyrme Model is an extreme case of the generalized one, where we neglect, as a first approximation, the $\Lag_2$ and $\Lag_4$ terms:

\begin{equation}
\Lag_{06} = \Lag_6 + \Lag_0.
\end{equation}

\noindent We can use the usual SU(2) parametrization of the Skyrme field $U$:

\begin{equation}
U=e^{i \xi \vec{n} \cdot \vec{\sigma}}= \cos \xi +i\sin\xi \vec n \cdot \vec \sigma \qquad \vec n^2 =1,
\end{equation}

\noindent where $\vec \sigma$ are the Pauli matrices, $\xi$ is a real field known as profile function, and $\vec n$ is an unit three component vector field related to a complex field $u$ by the stereographic projection

\begin{equation}
\vec{n}=\frac{1}{1+|u|^2} \left( u+\bar{u}, -i ( u-\bar{u}), 1- |u|^2 \right).
\end{equation}

\noindent Regarding this, and assuming that the potential only depends on $\Tr U$, we have

\begin{equation}
\Lag_{06}= \frac{  \lambda^2 \sin^4 \xi}{(1+|u|^2)^4} \;\left(  \epsilon^{\mu \nu \rho \sigma} \xi_{\nu} u_{\rho} \bar{u}_{\sigma} \right)^2
-\mu^2 \mathcal U(\xi).
\end{equation}

One of the main features of this BPS model is that it presents an infinite number of symmetries: the area-preserving diffeomorphisms on target space $S^2$ spanned by the complex field $u$ \cite{Symmetries}:

\begin{equation}
\xi \to \xi \, , \quad u\to \tilde u(u,\bar u, \xi),
\end{equation}

\noindent where 

\begin{equation}
(1+|\tilde u|^2)^{-2} d\xi d\tilde u d\bar{\tilde u} = (1+| u|^2)^{-2} d\xi  u d\bar{ u}.
\end{equation}

\noindent This symmetry is a subgroup of the full volume-preserving diffeomorphism (VPD) corresponding to the sextic term, however, it is the potential term depending on $\xi$ what breaks it. Furthermore, it exists another symmetry regarding the static energy functional which is invariant under VPD, but now on the base space.

Another important property of this model is the existence of a lower bound for the energy, what is called a BPS bound (BPS stands for Bogomolny-Prasad-Sommerfield), and it is here where the name of the model comes from. The idea behind getting the bound consists in trying to write the energy functional as a complete square, so finally we arrive at \cite{ASW}

\begin{eqnarray}
E &=& \int d^3 x \left(  \frac{\lambda^2 \sin^4 \xi}{(1+|u|^2)^4}  (\epsilon^{mnl} i \xi_m u_n\bar{u}_l)^2 +\mu^2 \mathcal U (\xi) \right) \nonumber \\
&\geq& 2\lambda \mu \pi^2 <\sqrt{\mathcal U}>_{S^3} |B|.
\end{eqnarray}

\noindent The saturation of the bound gives rise to the BPS equation

\begin{equation} \label{BPSeq}
\frac{\lambda \sin^2 \xi}{(1+|u|^2)^2} \epsilon^{mnl} i \xi_m u_n\bar{u}_l = \mp \mu \sqrt{\mathcal U},
\end{equation}

\noindent which implies to go from second order to first order equations.

To find some solutions to this BPS equation, first we have to choose the potential. For simplicity we take the standard Skyrme potential, although without the $\Lag_2$ term there is no relation to pion masses so another potential is possible:

\begin{equation}
\mathcal U=\frac{1}{2}\Tr (1-U) \;\; \rightarrow \;\; \mathcal U(\xi)=1- \cos \xi.
\end{equation}

\noindent The next step is to assume an ansatz. Here, although because of base space VPD's of static energy functional there are solutions with arbitrary shapes, we choose the simpler axially symmetric one:

\begin{equation} \label{ansatz}
\xi = \xi(r), \qquad  u(\theta, \phi)= \tan \frac{\theta}{2} e^{i n \phi},
\end{equation}

\noindent where $n$ is the baryon number. Then, it is easy to integrate the BPS equation getting an analytical compact solution with a radius proportional to $n^{1/3}$ and with a linear relation between mass (static energy) and baryon number:

\begin{equation}
E  = \frac{64\sqrt{2} \pi}{15} \mu \lambda |n|.
\end{equation}

In this sense, our model reproduces the phenomenological behaviour found in nuclei and encourages to further explore the nuclear world. Just as we will do in the next section with the study of binding energies, some nuclear thermodynamics and even neutron stars.

%%%%%%%%%%%%%%%%%%%%%%%%%%%%%%%%%%%%%%%%%%%
%%%%%%%%%%%%%%%%%%%%%%%%%%%%%%%%%%%%%%%%%%%

\section{Nuclear World}

\subsection{Binding Energies}
 The first thing we should do is to calculate the binding energies per nucleon to compare to experimental data.

\begin{equation}
E_{{\rm B},X} = Z E_{\rm p} + N E_{\rm n} - E_X,
\end{equation}

\noindent where $Z$ and $N$ are the number of protons and neutrons in a nucleus $X$ with static energy $E_X$, and $E_{\rm p}$ and $E_{\rm n}$ being the proton and neutron mass respectively (we are working wiht $c = 1$).

However, since the classical soliton energy is linear in baryon number, the binding energy is exactly zero. Therefore, we have to include further contributions to the energy in a natural way. More concretely we will introduce the spin and isospin quantization, the Coulomb energy and the isospin breaking \cite{ANSW1,ANSW2}:

\begin{equation}
E= E_{\rm sol} + E_{\rm SI} + E_{\rm C} + E_{\rm I}.
\end{equation}

\bigskip

 \noindent {\it Spin and isospin quantization} --- Spin and isospin are important since they are relevant quantum numbers of nuclei. The semi-classical quantization consists in introducing the spatial and isospin rotations around the classical solution, but with time-dependent coordinates parametrizing A and B SU(2) matrices:

\begin{equation}
U(t,\vec x)= {\rm A}(t)U_0(R_{\rm B}(t) \vec x){\rm A}^\dagger (t).
\end{equation}

\noindent Then, we plug it into the Lagrangian transforming the generalized velocities to the canonical momenta and the Lagrangian to the Hamiltonian, and get two copies of the symmetric top, one corresponding to spin and another to isospin. However, we have to distinguish the case $n = 1$ (proton and neutron) because the axial symmetry becomes spherical and spatial rotations are equivalent to isorotations, so the contribution we get is

\begin{equation}
E_{\rm SI} =  \frac{105}{512 \sqrt 2 \pi} \frac{3}{4} \frac{\hbar^2}{\lambda^2 \left( \frac{\mu}{\lambda} \right)^{1/3}}.
\end{equation}

\noindent And for nuclei with $n>1$:

\begin{eqnarray} 
E_{\rm SI} = \frac{105}{512 \sqrt 2 \pi} \frac{\hbar^2}{\lambda^2 \big( \frac{\mu}{\lambda n} \big)^{1/3}} \Bigl( \frac{ j (j+1)}{n^2} +\frac{4|i_3|(|i_3|+1)}{3n^2 +1} \Bigr), \nonumber
\end{eqnarray} 

\vspace{-.8cm}
\begin{equation}
\end{equation}

\noindent where $j$ is the spin and $i_3$ the third component of isospin.

\bigskip

\noindent {\it Coulomb energy} --- This contribution is important for high nuclei and is just the generalization of the usual expression for volume charge density

\begin{equation}
E_{\rm C}=\frac{1}{2 \varepsilon_0} \int d^3 x d^3 x' \frac{\rho(\vec r) \rho(\vec r\,')}{4 \pi|\vec r - \vec r\,'|}.
\end{equation}

Since this has a double integral, its calculation is too complicated. Then, to simplify things we can use the multipole expansion of the Coulomb potential with the decomposition of the charge density in spherical harmonics \cite{Marleau}. The result is, for $n = B = 1$:

\begin{equation} 
E_C^{\rm p} =\frac{1}{\sqrt{2} \pi\varepsilon_0} \Bigg( \frac{\mu}{\lambda } \Bigg)^{1/3} \Bigg(
\frac{128}{315 \pi^2}  + \frac{156625}{1317888} \Bigg) ,
\end{equation}

\begin{equation}
E_C^{\rm n} = \frac{1}{\sqrt{2} \pi\varepsilon_0} \Bigg( \frac{\mu}{\lambda } \Bigg)^{1/3} \Bigg(
\frac{128}{315 \pi^2}  - \frac{52585}{1317888} \Bigg),
\end{equation}

\noindent with different expressions for proton and neutron. And for $n = B > 1$

\begin{eqnarray}
&E_{\rm C}=& \frac{1}{\sqrt{2} \pi\varepsilon_0} \Bigg( \frac{\mu}{\lambda n} \Bigg)^{1/3}
\Bigg( 
\frac{128}{315 \pi^2 }  n^2 + \frac{245}{1536 }  n \; i_3 + \nonumber \\
&& + \frac{805}{5148 }  i_3^2 
 + \frac{7}{429 }  \frac{i_3^2}{(1+3n^2)^2} \Bigg).
\end{eqnarray}

\smallskip

{\it Isospin breaking} --- From the Coulomb energy we have that proton mass is heavier than neutron, but empirically we know is the other way around. This should be implemented in the effective Lagrangian with an isospin-breaking term, but it is obvious that the leading order contribution to the energy is

\begin{equation}
E_{\rm I} = a_{\rm I} i_3 \quad {\rm where} \quad a_{\rm I}<0 \quad \Leftrightarrow \quad M_{\rm n} >M_{\rm p}.
\end{equation}

\smallskip

Then, the idea now is to calculate numerical values for the masses of nuclei with our model so we can compare our binding energies per nucleon with experimental data. To do this, the first thing is to determine numerical values for the free parameters of our theory: $\mu$, $\lambda$ and $a_I$. Therefore, we fit them to the nuclear mass of proton, $M_p = 938.272 \; \MeV$, the neutron-proton mass difference, $M_n - M_p = 1.29333 \; \MeV$, and the nucleus with magical numbers Barium-138, $M(^{138}_{56} Ba) = 137.905 \; {\textrm u}$, where ${\textrm u}  = 931.494 \; \MeV$, getting the following values for them:

\vspace{-.7cm}
\begin{eqnarray}
\lambda \mu =   48.99 \; \MeV , \quad \left( \frac{\mu}{\lambda} \right)^\frac{1}{3} =  
0.6043 \; \textrm{fm}^{-1},  \nonumber
\end{eqnarray}
\begin{equation}
a_{\rm I} = -1.686 \; \MeV
\end{equation}

\begin{figure}[h]
\begin{center}
\includegraphics[width=0.47\textwidth]{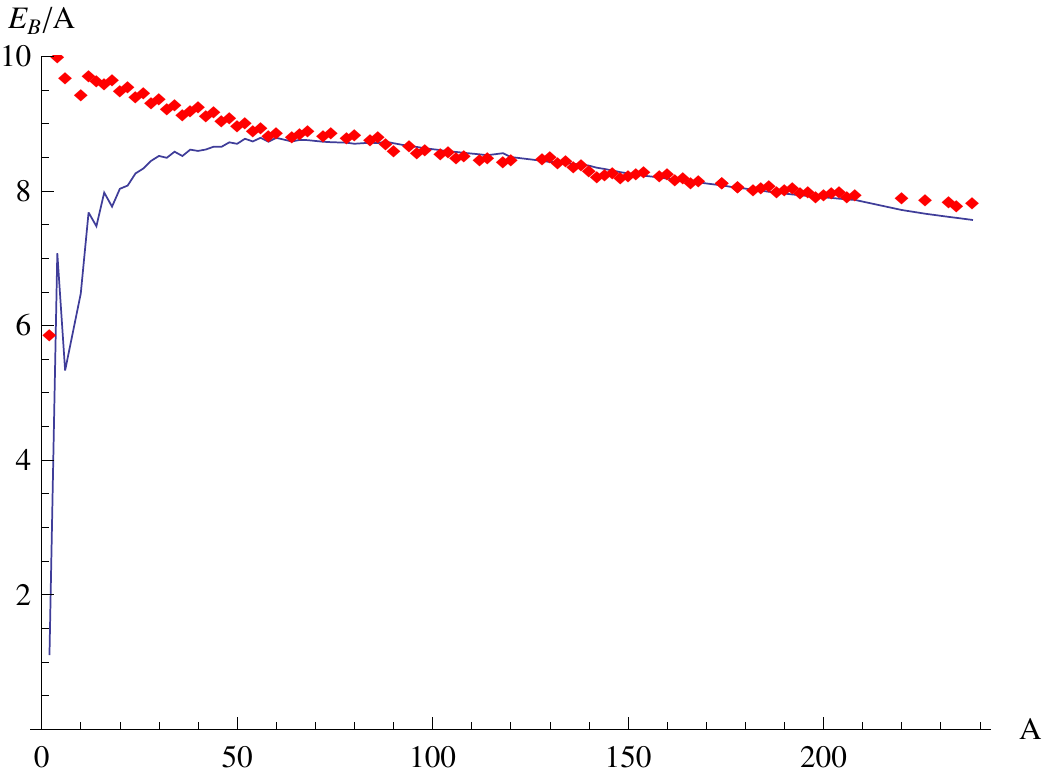} 
\caption{Binding energy per nucleon. Experimental values $\ldots$ solid line;  diamonds $\ldots $ results from our model.}
\label{FigEB}
\end{center}
\end{figure}

In comparing with experimental values, we follow the same strategy than in \cite{Marleau}: for each value of the atomic weight number $A$, we choose the values of $Z$ and $j$ corresponding to the most abundant nuclei, see fig. \ref{FigEB}.

Thus, we find an excellent agreement for sufficiently large nuclei whereas for small ones our model overestimates binding energies. This result is to be expected since in our Lagrangian we just have terms related to a collective behaviour. However, we expect this situation to improve with the inclusion of other terms, e. g., $\Lag_2$ and $\Lag_4$, and with solutions with different symmetries.

%%%%%%%%%%%%%%%%%%%%%%%%%%%%%%%%%%%%%%%%%%%

\subsection{Thermodynamics}

Now we are going to study a thermodynamical property of nuclear matter as it is the compressibility \cite{ANSSW}

\begin{equation}
\kappa = - \frac{1}{V} \left( \frac{dV}{dP} \right),
\end{equation}

\noindent a concept very similar to  the compression modulus defined by

\begin{equation}
\quad \mathcal{K} = \frac{9 V^2}{B} \frac{d^2 E}{d V^2}.
\end{equation}

In fact, if we have the standard thermodynamical relation for the pressure, $P = - \frac{dE}{dV}$, we can relate both definitions

\begin{equation} \label{relation}
\mathcal{K} = \frac{9 V}{B \kappa}.
\end{equation}

\noindent However, it is not obvious in our model since the definition of volume is not thermodynamical.

Again, one of the problems of the Standard Skyrme Model is that the compression modulus is too high. Then, as with the binding energies, we will try to improve this applying the BPS Skyrme Model. For this, the first thing we have to do is to introduce the pressure by calculating the energy-momentum tensor defined as

\begin{equation}
T^{\mu \nu} = - 2 \frac{\delta}{\delta g_{\mu \nu}} \int d^4 x \sqrt{|g|} \Lag_{06}.
\end{equation}

Thus, we arrive at a diagonal tensor with elements

\begin{eqnarray}
&& T^{00} = \lambda^2 \pi^4 \mathcal{B}_0^2 + \mu^2 \mathcal{U} = \mathcal{E}  \nonumber \\
&& T^{ij} = \delta^{ij} ( \lambda^2 \pi^4 \mathcal{B}_0^2 - \mu^2 \mathcal{U} ) = \delta^{ij} \mathcal{P},
\end{eqnarray}

\noindent where here $\mathcal B_0$ is the zero-component of the baryon current given in the equation (\ref{Bmu}). On the other hand, the spacial component of the tensor, $T^{ij}$ is just proportional to the BPS equation (\ref{BPSeq}), but without assuming any ansatz yet. Therefore, it is trivial that for BPS solutions the pressure is exactly zero. And even more, the conservation equation implies that any static solution has constant pressure:

\begin{equation}
\partial_\mu T^{\mu \nu} = 0 \quad \Rightarrow \quad \mathcal{P} = P = const.
\end{equation}

Now, we can translate this constant pressure equation to the spherically symmetric ansatz (\ref{ansatz}):

\begin{equation}
\frac{|B| \lambda}{2 r^2} \sin^2 \xi \xi_r = - \mu \sqrt{\mathcal U + \tilde P} \qquad (P = \mu^2 \tilde P ),
\end{equation}

\noindent where $B$ is the baryon number and $\tilde P = P/\mu^2$. Introducing the new coordinate $z$ and field $\eta$:

\begin{equation}
z = \frac{2 \mu}{3 |B| \lambda} r^3, \qquad  \eta = \frac{1}{2} \left( \xi - \frac{1}{2} \sin 2 \xi \right),
\end{equation}

\noindent the equation can be written as

\begin{equation}
\eta_z = - \sqrt{\mathcal{U} + \tilde{P}},
\end{equation}

\noindent and by direct integration (the volume is just the value of $z$ for which the profile field goes to zero, multiplied by some constants) we arrive at the equation of state 

\begin{eqnarray}
&V(P)& =  2 \pi |B| \frac{\lambda}{\mu} \int^\pi_0 \frac{\sin ^2 \xi d \xi}{\sqrt{\mathcal U + \tilde P}} \nonumber \\
\smallskip
&& = 2 \pi |B| \frac{\lambda}{\mu} \int^{\frac{\pi}{2}}_0 \frac{d \eta}{\sqrt{\mathcal U + \tilde P}}.
\end{eqnarray}

We can also calculate the expression for the energy with the help of the constant pressure equation getting

\begin{eqnarray}
&E(P)& =2 \pi \lambda \mu |B| \int^\pi_0 d \xi  \sin^2 \xi \frac{2 \mathcal U + \tilde P}{\sqrt{\mathcal U + \tilde P}} \nonumber \\
&& = 2 \pi \lambda \mu |B| \int^{\frac{\pi}{2}}_0 d \eta \frac{2 \mathcal U + \tilde P}{\sqrt{\mathcal U + \tilde P}}.
\end{eqnarray}

\noindent Then, from these two expressions it is easy to see that the standard thermodynamical relation for the pressure holds, and therefore the relation (\ref{relation}) between $\kappa$ and $\mathcal K$ also applies in our model. But this is not the only nice property of the BPS model, as we have pointed before, one of its advantages is the possibility of analytical calculations. We will see here an example with the potential

\begin{equation}
\mathcal U = \eta^{2/3},
\end{equation}

\noindent corresponding to a quadratic potential in the original field $\xi$. Then, the non-zero pressure equation is

\begin{equation}
\eta_z = - \sqrt{\eta^{2/3} + \tilde P},
\end{equation}

\noindent with implicit solution

\begin{eqnarray}
&&\frac{3}{2} \Bigg[ \sqrt{\eta^{2/3} + \tilde P} \eta^{1/3}  \nonumber \\
&& - \tilde P \ln \left( 2 \left( \sqrt{\eta^{2/3} + \tilde P} + \eta^{1/3} \right) \right) \Bigg] = z_0 - z .
\end{eqnarray}

Finally, the corresponding volume  ($V = 2 \pi |B| \frac{\lambda}{\mu} \tilde V$) is

\begin{eqnarray}
\tilde V (\tilde P)&=& \frac{3}{2} \Bigg[ \sqrt{\left( \frac{\pi}{2} \right)^{2/3} + \tilde P} \left( \frac{\pi}{2} \right)^{1/3}  \nonumber \\
&& - \tilde P \ln \left( \sqrt{\left( \frac{\pi}{2} \right)^{2/3} + \tilde P} + \left( \frac{\pi}{2} \right)^{1/3} \right) \Bigg] \nonumber \\
&& + \frac{1}{2} \tilde P \ln \tilde P .
\end{eqnarray}

\noindent It is easy to see that the term $\tilde P \ln \tilde P$ leads to  an infinite compressibility, i.e., to a zero compression modulus. This does not mean it costs zero energy to squeeze a BPS Skyrmion under external pressure, but that the pressure used to squeeze the soliton and the resulting small change in volume are not linearly related.

\begin{figure}[h]
\begin{center}
\includegraphics[width=0.47\textwidth]{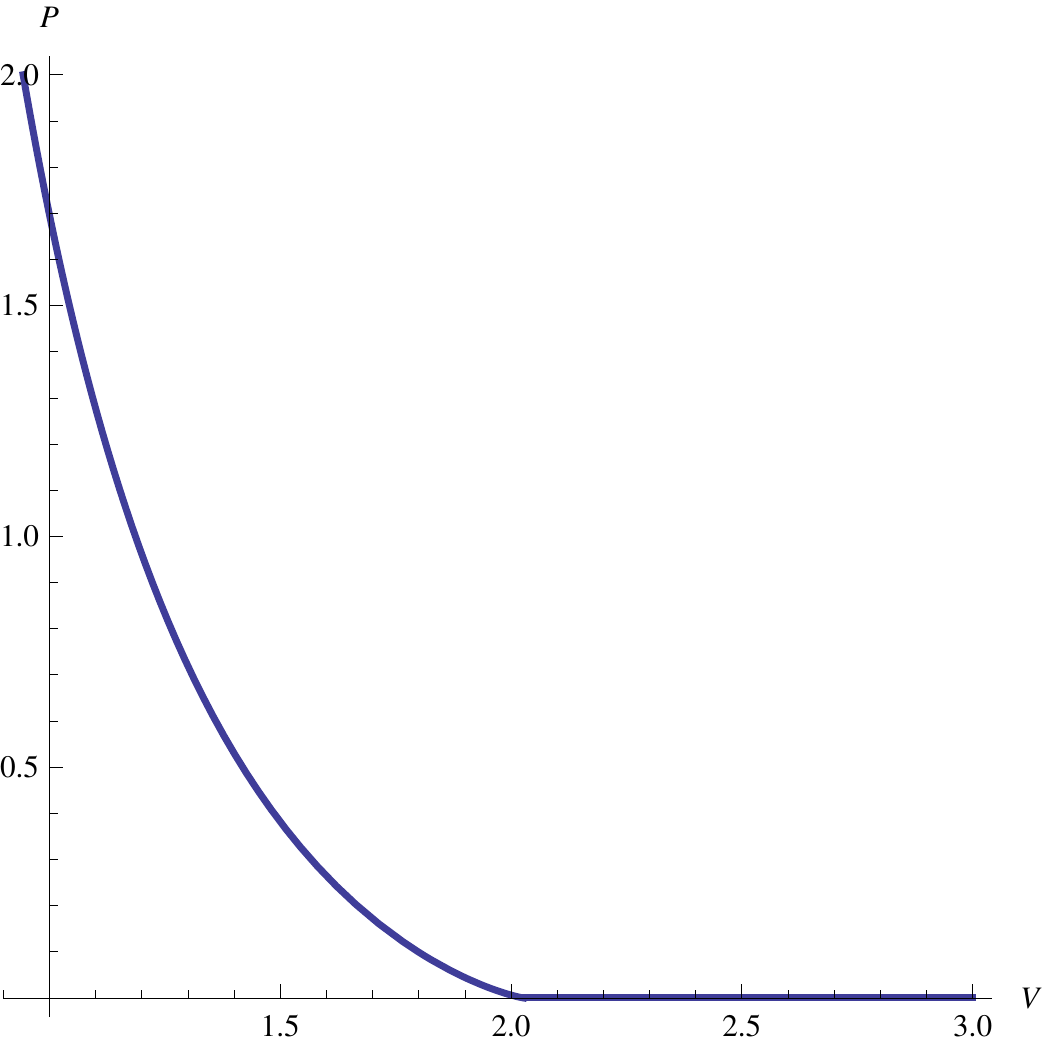} 
\caption{Equation of state for the potential $\mathcal U = \eta^{2/3}$.}
\label{eqState}
\end{center}
\end{figure}

Finally, in fig. \ref{eqState} we present the corresponding equation of state. We see that for a volume $V_0$ there is a phase transition between an ideal gas of non-overlapping compactons with zero pressure for $V>V_0$, and a kind of liquid phase for $V<V_0$ (see \cite{ANSSW} for details, these results are rather generic and hold for many potentials).

%%%%%%%%%%%%%%%%%%%%%%%%%%%%%%%%%%%%%%%%%%%

\subsection{BPS Neutron Stars}

The final topic in this nuclear world is devoted to one important issue under current research, the coupling of the Skyrme model, and more concretely its BPS version, to gravity and the study of neutron stars \cite{ANSVW}. To do this, we just have to recall Einstein's equations:

\begin{equation}
G_{\mu \nu} = \frac{\kappa^2}{2} T_{\mu \nu},
\end{equation}

\noindent where $\kappa^2 = 16 \pi G = 6.654 \cdot 10^{-41} \; \fm \; \MeV^{-1}$. Again, we will use an axially symmetric ansatz for the Skyrme field, but here, we have to introduce a static, spherically symmetric metric too:

\begin{eqnarray}
ds^2 = {\bf A}(r) dt^2 - {\bf B}(r) dr^2 -r^2 (d \theta^2 + \sin^2 \theta d \phi^2). \nonumber \\
&&
\end{eqnarray}

Then, the only thing we need is to calculate the Einstein tensor, $G_{\mu \nu}$, and the energy-momentum tensor, $T_{\mu \nu}$, both diagonal tensors because of the symmetry of our system. Defining a new variable $h$:

\begin{equation}
h = \frac{1}{2} (1 - \cos \xi),
\end{equation}

\noindent we finally arrive at the Einstein equations ($' \equiv \partial_r$)

\begin{eqnarray}
\frac{1}{r} \frac{{\bf B}'}{{\bf B}} &=& -\frac{1}{r^2} ({\bf B}-1) + \frac{\kappa^2}{2} {\bf B} \rho, \\
r ({\bf B} p)' &=& \frac{1}{2} (1 - {\bf B}) {\bf B} (\rho + 3p) \nonumber \\
&&+ \frac{\kappa^2}{2} \mu^2 r^2 {\bf B}^2 \mathcal{U}(h)p, \\
\frac{{\bf A}'}{{\bf A}} &=& \frac{1}{r} ({\bf B} - 1) + \frac{\kappa^2}{2} r {\bf B} p,
\end{eqnarray}

\noindent which are a system of two ODEs for $h$ and ${\bf B}$, plus a third equation determinig ${\bf A}$ in terms of $h$ and ${\bf B}$. Further, $\mathcal U(h)$ is the potential coming from the $\Lag_0$ term in the Lagrangian, and $\rho$ and $p$ are the energy density and pressure respectively, which for the axially symmetric ansatz read

\begin{equation}
\rho = \frac{4 B^2 \lambda^2}{{\bf B} r^4} h (1-h) h_r^2 + \mu^2 \mathcal U(h),
\end{equation}

\begin{equation}
p = \rho - 2 \mu^2 \mathcal U(h).
\end{equation}

Thus, the next step is to study the solutions of these equations describing neutron stars. For this, we solve them numerically by a shooting from the center. The boundary conditions we impose are $h(r=0) = 1$ and ${\bf B} (r=0) = 1$, and expanding about the center we see we only have one free paremeter left, $h_2$: $h(r) \sim 1 - (1/2) h_2 r^2 + \cdots$ (see \cite{ANSVW} for more details). Here, we will consider two different potentials: the pion mass potential $\mathcal U_\pi = 1 - \cos \xi$, and its square, $\mathcal U_4 = \mathcal U_\pi^2$, which presents a quartic behaviour near the vacuum. We need to give some values for the model parameters ${\bf m} = \lambda \mu$ and ${\bf l} = (\lambda/\mu)^{1/3}$, which for these potentials will be \cite{ANSVW}

\begin{eqnarray}
&\mathcal U_\pi: & \; {\bf m} = 49.15 \; \MeV, \quad {\bf l} = 0.884 \; \fm \\
&\mathcal U_\pi^2: & \; {\bf m} = 47.20 \; \MeV, \quad {\bf l} = 0.746 \; \fm
\end{eqnarray}

It is worth to comment that for these potentials we have compacton solutions with a radius $R$. Then, for $r \geq R$, $h(r) = 0$ and ${\bf B} (r) = (1 - \frac{2 GM}{r})^{-1}$, where $M$ is the physical mass of the neutron star.

The main result we get here is the maximum value of the baryon number, $B_{max}$, for which solutions still exist with the corresponding mass and radius of this neutron star. It is convenient to give $B$ in term of solar mass units, so we will use $n \equiv (B/B_\odot)$ instead of $B$. Then,

\begin{eqnarray}
&\mathcal U_\pi: & \; n_{max} = 5.005, \quad M_{max} = 3.734 M_\odot, \nonumber \\
&& R_{max} = 18.458 \; \textrm{km}, \\
&\mathcal U_\pi^2: & \; n_{max} = 3.271, \quad M_{max} = 2.4388 M_\odot, \nonumber \\
&& R_{max} = 16.801 \; \textrm{km}.
\end{eqnarray}

\begin{figure}[h]
\begin{center}
\includegraphics[width=0.47\textwidth]{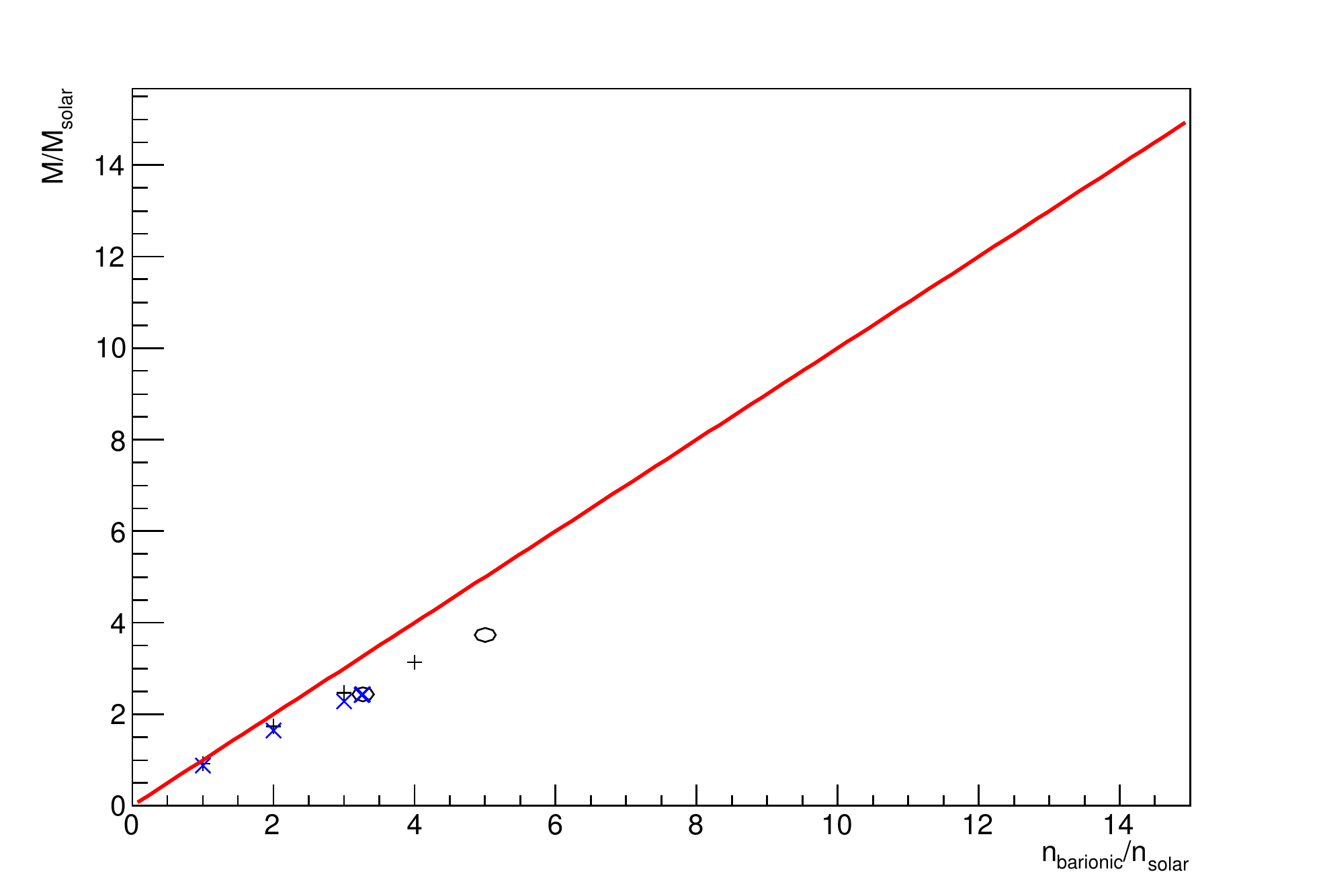} 
\caption{Neutron star mass as function of baryon number, both in solar units. Symbol plus ($+$): potential $\mathcal U_\pi$. Symbol cross ($\times$): potential $\mathcal U_\pi^2$.}
\label{Mn}
\end{center}
\end{figure}

\begin{figure}[h]
\begin{center}
\includegraphics[width=0.47\textwidth]{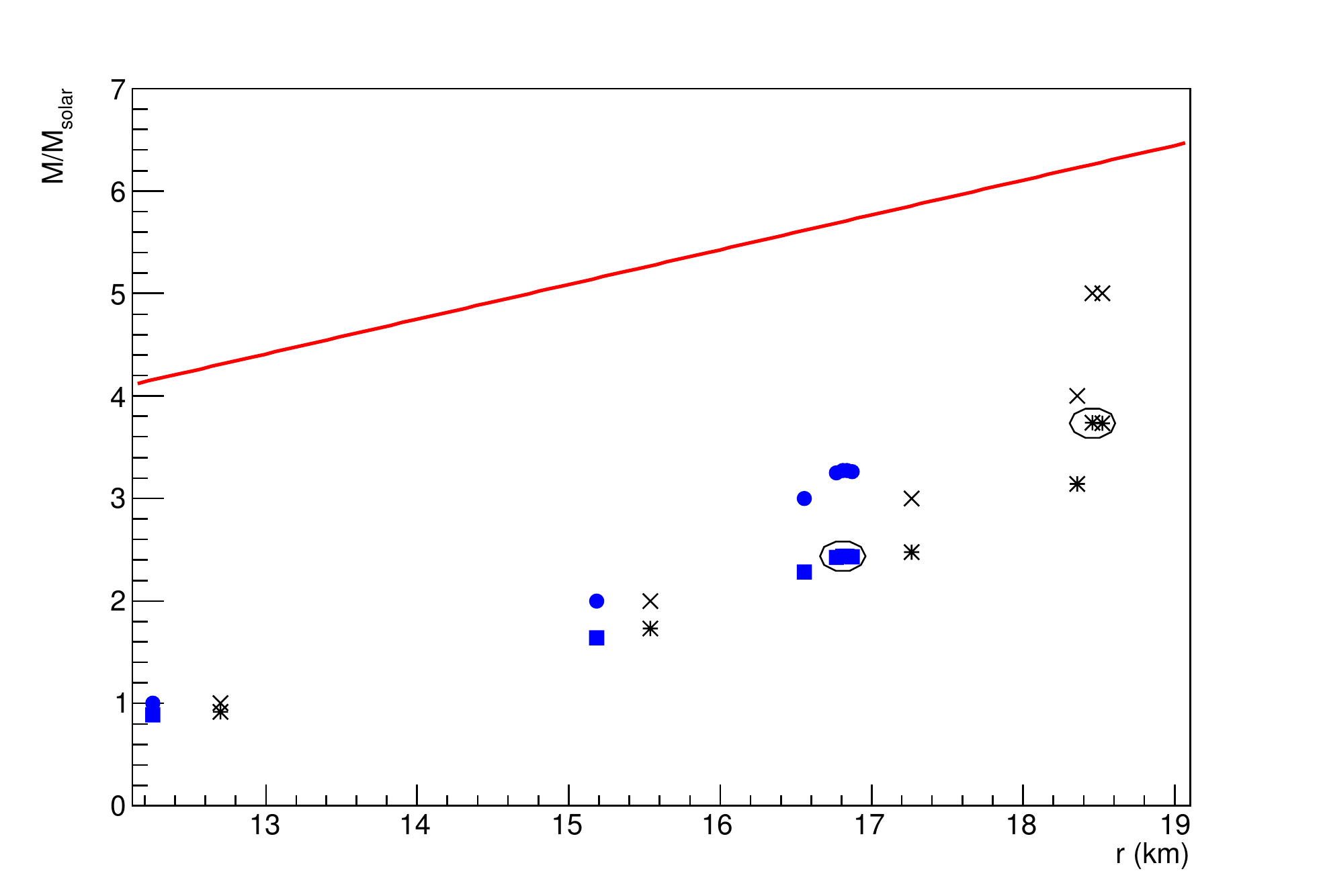} 
\caption{Neutron star mass as a function of the neutron star radius. Symbol asterisk ($\ast$): potential $\mathcal U_\pi$. Symbol square: potential $\mathcal U_\pi^2$. Maximum values are indicated by circles. Also the value of $n$ is plotted. The cross ($\times$) corresponds to $\mathcal U_\pi$ and the circle to $\mathcal U_\pi^2$.}
\label{MR}
\end{center}
\end{figure}

\noindent It is well know that there are neutron stars with masses up to about $M \sim M_\odot$, as well as indications for masses up to about $2.5 M_\odot$ \cite{NeutronStar1,NeutronStar2}, whereas the radius are expected to be in a range $R \sim 10-20$ km. Therefore, we got results in excellent agreement with the observations and this subject becomes a promising research field. In figures \ref{Mn} and \ref{MR} we can see these results for $n_{max}$ and other different values of $n$. In \ref{Mn} we have the mass as function of $n$, and in \ref{MR} we present the mass-radius relation.

%%%%%%%%%%%%%%%%%%%%%%%%%%%%%%%%%%%%%%%%%%%
%%%%%%%%%%%%%%%%%%%%%%%%%%%%%%%%%%%%%%%%%%%

\section{Conclusions and outlook}

A first point we want to emphasize is the existence of a novel Skyrme model as a limit of generalized ones with analytical solutions, the so-called BPS Skyrme Model. Because of its BPS property, it shows a linear mass-baryon number relation which agrees with the phenomenological behaviour of nuclei, so it becomes a good starting point for the study of nuclei and nuclear matter. With this in mind we have studied the binding energies of nuclei. For this we need to take into account further but natural small contributions to the energy coming from the semiclassical quantization (introducing the spin and isospin), Coulomb energy and isospin-breaking. Then, we find that this is a good approximation for heavy nuclei although it overestimates the binding energy of lighter ones. However, there is work in progress on this issue and it is expected to improve with the inclusion of additional terms to the Lagrangian and the use of other shapes instead of the axially symmetric ansatz.

Further, we have extended this study to nuclear matter and neutron stars both within the BPS Skyrme Model. Thus, we found that in the case of the compression modulus of nuclear matter where the Standard Skyrme model gave high values, now we classically get a zero one. A more realistic treatment certainly requires the quantization of some vibrational modes. In any case, this means that now pressure and volume are not linearly related. 

And finally, we have started onother promising field of study, namely the case of neutron stars. Here, after coupling the BPS model to gravity we found and solved the Einstein equations resulting in numerical results in perfect agreement with physical data both for masses and radii.

There are some obvious directions of further research. On the one hand, as said, this BPS model is a kind of approximation for a more general one where other small contributions are included, e.g, $\Lag_2$. We then want to study the effect of adding these new terms. For instance, with this $\Lag_2$ the infinite VPD symmetry no longer holds, so the ``right'' shape for each topological sector should results.

On the other hand, related to neutron stars, one first step is to study the effect of different potentials to see which give a more realistic description. Further, we want to get the corresponding equation of state (equation relating pressure and energy density). Our first results show that it is possible but with a novel and really interesting outcome: the equation of state of neutron stars depends on the baryon number, $B$, and the gravitational coupling constant, $\kappa$.

%%%%%%%%%%%%%%%%%%%%%%%%%%%%%%%%%%%%%%%%%%%
%%%%%%%%%%%%%%%%%%%%%%%%%%%%%%%%%%%%%%%%%%%

\section*{Acknowledgement}
The authors acknowledge financial support from the Ministry of Education, Culture and Sports, Spain (grant FPA2011-22776), the Xunta de Galicia (grant INCITE09.296.035PR and Conselleria de Educacion), the Spanish Consolider-Ingenio 2010 Programme CPAN (CSD2007-00042), FEDER and the UK Engineering and Physical Sciences Research Council. CN thanks the Spanish Ministery of Education, Culture and Sports for financial support (grant FPU AP2010-5772).



%% References

\nocite{*}
\bibliographystyle{elsarticle-num}
\bibliography{martin}



\end{document}